\documentclass[iop,apjl]{emulateapj}
\usepackage{amsmath}
\usepackage{graphicx}
\usepackage{natbib}
\usepackage{color}
\usepackage[colorlinks,citecolor=blue,linkcolor=red]{hyperref}
\usepackage{threeparttable}
\usepackage{rotating}
\usepackage{multirow,booktabs}
\usepackage{wasysym}
\usepackage{float} 
  
\newcommand{\lum}{erg~s\ensuremath{^{-1}}}

\begin{document}
\bibliographystyle{apj} 
\title{Mid-Infrared Variability of Changing-look AGNs}

\author{Zhenfeng Sheng\altaffilmark{1,2$\dag$}, Tinggui Wang\altaffilmark{1,2$\ddag$}, Ning Jiang\altaffilmark{1,2}, Chenwei Yang\altaffilmark{1,2}, Lin Yan\altaffilmark{3,4}, Liming Dou\altaffilmark{5,6}, Bo Peng\altaffilmark{1,2}}

\altaffiltext{1}{CAS Key Laboratory for Researches in Galaxies and Cosmology, University of Sciences and Technology of China, Hefei, Anhui 230026, China; $^{\dag}$shengzf@mail.ustc.edu.cn; $^{\ddag}$twang@ustc.edu.cn}

\altaffiltext{2}{School of Astronomy and Space Science, University of Science and Technology of China, Hefei 230026, China;}
\altaffiltext{3}{Caltech Optical Observatories, Cahill Center for Astronomy and Astrophysics, California Institute of Technology, Pasadena, CA 91125, USA} 
\altaffiltext{4}{Infrared Processing and Analysis Center, California Institute of Technology, Pasadena, CA 91125, USA}
\altaffiltext{5}{Center for Astrophysics, Guangzhou University, Guangzhou
510006, China} 
\altaffiltext{6}{Astronomy Science and Technology Research
Laboratory of Department of Education of Guangdong Province, Guangzhou 510006, China}

\begin{abstract}
It is known that some active galactic nuclei (AGNs) transit from Type 1 to Type 2 or vice versa. There are two explanations for the so-called changing-look AGNs: one is the dramatic change of the obscuration along the line of sight, and the other is the variation of accretion rate. In this Letter, we report the detection of large amplitude variations in the mid-infrared luminosity during the transitions in 10 changing-look AGNs using \emph{Wide-field Infrared Survey Explorer (WISE)} and \emph{newly released Near-Earth Object WISE Reactivation} data. The mid-infrared light curves of 10 objects echo the variability in the optical band with a time lag expected for dust reprocessing. The large variability amplitude is inconsistent with the scenario of varying obscuration, rather it supports the scheme of dramatic change in the accretion rate.

\end{abstract}

\section{Introduction}

Active galactic nuclei (AGNs) are empirically classified into Type 1 and Type 2 according to the emission line widths. Type 1 AGNs show both broad and narrow emission lines in spectra while Type 2 display only narrow lines. Intermediate types, 1.5 and 1.8/1.9, were further introduced depending on the relative strength of broad and narrow lines (Osterbrock et al.~\citeyear{Osterbrock1977},~\citeyear{Osterbrock1981}). Whereas the early discovery of broad lines in the polarized spectra of Type 2 AGNs (Antonucci \& Miller,~\citeyear{Anton1985}), together with other evidence, led to a unification scheme (Antonucci et al. \citeyear{Antonucci1993}): two types of AGNs are intrinsically the same but differ only in the orientation of the torus-like obscurer. In this scheme, Type 1 AGNs are viewed face-on so that we look directly into the central accretion disk and the broad emission line region (BLR), while Type 2 AGNs are viewed edge-on and our line of sight to the central engine is blocked by a putative dusty torus. Despite the success of the unification model, there are arguments that at least some type-2 AGNs are intrinsic, lacking of broad lines because of an inadequate accretion rate (Shi et al. \citeyear{shi2010}; Bianchi et al. \citeyear{Bianchi2012}; Pons et al. \citeyear{Pons2014}). 

Some AGNs are known to transit between Type 1 and Type 2 (e.g. from Type 2 to Type 1, Khachikian $\&$ Weedman \citeyear{Khachikian1971}; from Type 1 to Type 2, Peterson \& Perez \citeyear{Peterson1984}). These objects are called changing-look (CL) AGNs, featuring emerging or disappearing broad emission lines (BELs). There are some notable CL AGNs reported so far. Mrk 590 changed from Seyfert 1.5 to 1.0 and back to 2 over several decades (Denney et al. \citeyear{Denney2014}). NGC 2617 was reported to have changed from Type 1.8 to Type 1 (Shappee et al. \citeyear{Shappee2014}), but recently it likely has a new outburst and continues brightening (Oknyansky et al. \citeyear{Oknyansky2017}). More recently, it has been reported that Mrk 1018 changed back to Type 1.9 after 30 years of being a as Type 1 (McElroy et al. \citeyear{McElroy2016}).  
\begin{table*}
	\caption{\label{t1}Information of 24 Changing-look AGNs}
	\centering
	\begin{tabular}{cccccccccc}
		\toprule
		(1)&(2)&(3)&(4)&(5)&(6)&(7)&(8)&(9)\\
		Name & $t_{spec}$  & $\Delta t_{wise-spec}$ & Max $\Delta$\emph{W1} & Max $\Delta$\emph{W2} & $\sigma_{W1}$ & $\sigma_{W2}$ & Transition & note\\
		& (MJD)  & (year)& (mag)& (mag)\\
		\midrule
		J002311.06+003517.5 & 51816, 55480 & -0.29 & 0.47$\pm$0.03 & 0.28$\pm$0.04& 13.62 & 6.89 & A BELs     & Macleod et al. \citeyearpar{MacLeod2016}\\
		J015957.64+003310.4 & 51871, 55201 & 0.03  & 0.21$\pm$0.06 & 0.26$\pm$0.12 & 3.67 & 2.28 & D BELs  & LaMassa et al. \citeyearpar{LaMassa2015}\\
		J012648.08-083948.0 & 52163, 54465 & 2.53 &  0.06$\pm$0.04 & 0.16$\pm$0.08 & 1.72 & 1.87 & D BELs  & Ruan et al. \citeyearpar{Ruan2015} \\
		J022556.07+003026.7 & 52944, 55445 & -0.11 & 0.23$\pm$0.04 & 0.39$\pm$0.12 & 5.31 & 3.30& D \& A BELs	        & Macleod et al. \citeyearpar{MacLeod2016}\\
		J022652.24 -003916.5 & 52641, 56267 & -2.88 & 0.23$\pm$0.06 & 0.48$\pm$0.14 & 3.53 & 3.53 & D BELs  & Macleod et al. \citeyearpar{MacLeod2016}\\
		J035301.02 -062326.3 & 51908, 54853 &  1.05  & 0.18$\pm$0.02 & 0.24$\pm$0.03 & 10.36 & 9.14 & 1.8 $\rightarrow$ 1 & Runco et al. \citeyearpar{Runco2016}\\ 
		J081319.34+460849.5 & 51877, 55210 &  0.24 & 0.33$\pm$0.01 & 0.57$\pm$0.02 & 27.47 & 31.42 & 1.8 $\rightarrow$ 1 & Runco et al. \citeyearpar{Runco2016}\\ 
		J084748.28+182439.9 & 53711, 54852 &  1.26 & 0.29$\pm$0.01 & 0.34$\pm$0.02 & 22.19 & 14.93 & 1 $\rightarrow$ 1.9 $\rightarrow$ 2 & Runco et al. \citeyearpar{Runco2016}\\
		J090902.35+133019.4 & 53826, 55210 &  0.29 & 0.70$\pm$0.02 & 1.20$\pm$0.05 & 41.98 & 24.57 & 1.8 $\rightarrow$ 1 & Runco et al. \citeyearpar{Runco2016}\\
		J093812.27+074340.0 & 52733, 55210 &  0.32 & 0.04$\pm$0.01 & 0.07$\pm$0.01 & 4.26 & 4.76 & 1 $\rightarrow$ 1.8 & Runco et al. \citeyearpar{Runco2016}\\
		J094838.43+403043.5 & 52709, 55211 &  0.29 & 0.19$\pm$0.01 & 0.16$\pm$0.01 & 20.94 & 15.41 & 1 $\rightarrow$ 1.8 & Runco et al. \citeyearpar{Runco2016}\\
		J100220.17+450927.3 & 52376, 56683 & -3.74 & 0.30$\pm$0.03 & 0.26$\pm$0.06 & 9.36 & 4.58 &  D BELs  & Macleod et al. \citeyearpar{MacLeod2016}\\
		J101152.98+544206.4 & 52652,57073  & -4.82 &1.17$\pm$0.04 & 1.76$\pm$0.08 & 33.19 & 22.69 & D BELs& Runnoe et al. \citeyear{Runnoe2016}\\
		J102152.34+464515.6 & 52614, 56769 & -3.97 & 0.65$\pm$0.03 & 0.73$\pm$0.03 & 25.68 & 21.27 & D BELs  & Macleod et al. \citeyearpar{MacLeod2016}\\
		J132457.29+480241.2 & 52759, 56805 & -3.97 & 0.45$\pm$0.02  & 0.47$\pm$0.02 & 27.55 & 17.65 & D BELs  & Macleod et al. \citeyearpar{MacLeod2016}\\
		J154507.53+170951.1 & 53889, 54936 &  0.82 & 0.53$\pm$0.01 & 0.66$\pm$0.01 & 49.72 & 47.89 & 1.8 $\rightarrow$ 1 & Runco et al. \citeyearpar{Runco2016}\\
		J155440.25+362952.0 & 53172,57543 & -6.34 & 0.67$\pm$0.02 & 0.96$\pm$0.03 & 35.93 & 27.76 & 2$\rightarrow$1 & Gezari et al. \citeyearpar{Gezari2017}\\
		J214613.31+000930.8 & 52968, 55478 & -0.39 & 0.14$\pm$0.04 & 0.13$\pm$0.06 & 3.48 & 2.20 & A BELs     & Macleod et al. \citeyearpar{MacLeod2016}\\
		J225240.37+010958.7 & 52174, 55500 & -0.40 & 0.65$\pm$0.06 & 0.88$\pm$0.09 & 11.26 & 9.93 & A BELs     & Macleod et al. \citeyearpar{MacLeod2016}\\
		J233317.38 -002303.4 & 52199, 55447 & -0.23 & 0.28$\pm$0.04 & 0.14$\pm$0.05 & 7.45 & 2.81 & A BELs     & Macleod et al. \citeyearpar{MacLeod2016}\\
		J233602.98+001728.7 & 52096, 55449 & -0.23 & 0.40$\pm$0.08 & 0.71$\pm$0.25 & 5.20 & 2.78 & D BELs & Ruan et al. \citeyear{Ruan2015}\\
		Mrk 590 & 52649, 56664 & -3.97 & 0.13$\pm$0.01 & 0.34$\pm$0.01 & 18.61 & 43.95 & 1.5 $\rightarrow$ 1 $\rightarrow$ 2 & Denney et al. \citeyearpar{Denney2014}\\
		Mrk 1018 & 51812, 57033 & -4.99 & 0.76$\pm$0.01 & 1.05$\pm$0.01 & 79.92 & 105.78 & 1.9 $\rightarrow$ 1 $\rightarrow$ 1.9 & McElroy et al. \citeyearpar{McElroy2016}\\
		NGC 2617 & 53003, 56407 & -2.99 & 0.64$\pm$0.01 & 0.87$\pm$0.01 & 74.38 & 93.95 & 1.8 $\rightarrow$ 1 & Shappee et al. \citeyearpar{Shappee2014}\\
		\bottomrule
	\end{tabular}
	\tablecomments{The information of 24 CL AGNs. $t_{spec}$ lists the MJD of two spectra be used to confirm the type transition; $\Delta t_{wise-spec}$ lists the interval between first \emph{WISE} data point and the spectrum epoch ( second MJD in $t_{spec}$ ) that confirmed transition ($\Delta t_{wise-spec}<0$ means that the transition is more likely to to be covered by \emph{WISE/NEOWISE}); columns 4 and 5 list the maximum variation of the \emph{W1} and \emph{W2} bands, while columns 6 and 7 are the corresponding variation significance of \emph{W1} and \emph{W2}; column 8 lists the transition of each source, A BELs means appear, while D BELs means disappear; last column lists the corresponding reference}
\end{table*}

Although the origin of the CL behavior is not well understood, various scenarios have been proposed. In one scenario, CL is interpreted in the context of the unification scheme, and disappearing or emerging of BELs are ascribed to the variable obscurer moving in and out of the line of sight (Goodrich \citeyear{Goodrich1989}). In the other scenario, CL is attributed to the dramatic changes in accretion rate, arising from disk instability or even the tidal disruption, in which the continuum and broad lines should respond immediately while narrow lines remain nearly unchanged. LaMassa et al. \citeyearpar{LaMassa2015} demonstrated that both the photometric and spectral properties of CL AGN J0159+0033 cannot be explained by the unification paradigm, but suggested that accretion power decreases. Macleod et al. (\citeyear{MacLeod2016}) undertook a systematic search for CL AGNs using SDSS and Pan-STARRS1 and found 10 of them. Runnoe et al. (\citeyear{Runnoe2016}) recently reported a new CL AGN, J1011+5442, through the Time Domain Spectroscopic Survey. Both Macleod et al. (\citeyear{MacLeod2016}) and Runnoe et al. (\citeyear{Runnoe2016}) favor the interpretation of accretion rate change. 
 
In this work, we focus on the mid-infrared variability (MIR) of CL AGNs and its application in testing CL scenarios. Because the infrared emission is produced by dust heated by the UV radiation of accretion disk, it would respond to the variation of the latter with a time lag of order of years. Jun et al. (\citeyear{Jun2015}) used the mid-infrared echo to confirm that PG 1302-102's optical periodic variability is accretion disk driven. Besides, infrared emission is much less affected by dust extinction than optical radiation as the opacity decreases steeply toward long wavelength. Moreover, the size of the torus is much larger than those of the BLR and accretion disk. Therefore, the effect of obscuration by a dusty cloud in the optical and infrared would be very different. This would allow us to test the two different scenarios. 

We report a discovery of a significant infrared variation of eight CL AGNs. The outline of this Letter is as follows. In \S\ref{Dt}, we describe the Catalina Real-Time Transient
Survey (CRTS) data, \emph{WISE/WISE/Near-Earth Object WISE Reactivation (NEOWISE)} data, and Sloan Digital Sky Survey (SDSS) data used in this study, along with initial data processing. In \S\ref{lc}, we present some details of each source and shortly review their properties. In \S\ref{disc}, we simply discuss the possible scenarios, and then we come to a conclusion in \S\ref{C}. We adopt a flat $\Lambda CDM$ cosmology with $H_0=70 kms^{-1} Mpc^{-1}$ and $\Omega_m=0.27$.  

\begin{figure*}
	\figurenum{1}
	%\epsscale{0.95}
	\plotone{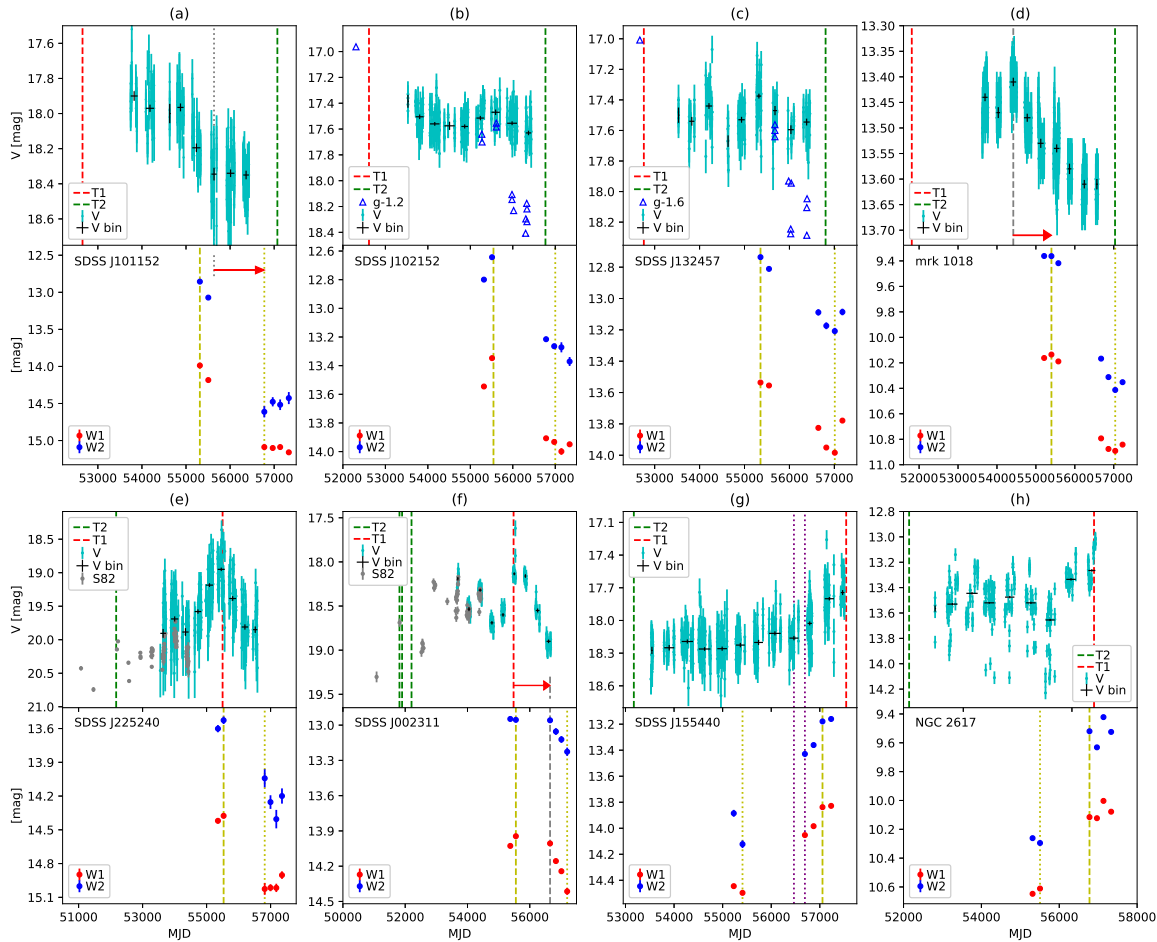}
	\caption{Changing-look AGNs with disappearing BELs:(a) J101152, (b) J102152, (c) J132457, and (d) Mrk 1018; AGNs with emerging BELs: (e) J225240, (f) J002311, (g) J155440, and (h) NGC 2617. In each top panel,the cyan dots with error bars are \emph{V}-band data from CRTS. The black crosses represent the median value of each season epoch, while the y-error bars are calculated by error propagation method. The red dashed line marks the epoch of the spectrum used to confirm Type 1 (T1) while the green dashed line marks the epoch of spectrum used confirm Type 2 (T2). For J102152 and J132457, the blue open triangles are \emph{g}-band data from MacLeod et al (\citeyear{MacLeod2016}). The gray dots represent the \emph{g}-band data of Strip 82 (Abazajian et al. \citeyear{Abazajian2009}) with a constant offset $m_0$. In each bottom panel, the red and blue dots represent the median value of \emph{W1} and \emph{W2} bands, respectively, while the corresponding error bars are also the propagation error. The yellow dashed/dotted lines not only mark the bright/dim state of each source, respectively, but also mark the start/end of the significant MIR variation time scale. The red arrow marks the upper limit lag shift between optical and MIR light curves. For J155440, the purple dotted lines mark the time range when the "turn on" event occurred. J225240 is very dim in the \emph{V}-band, so we constrain the CRTS measurement error $\leq$0.4 when performing data processing.}
	\label{fig:1}
\end{figure*} 

\section{Data}\label{Dt}
In this section, we will introduce the datasets used to construct the optical/MIR light curves and the sample selection. Our investigation is mainly based on \emph{V}-band data from CRTS (Drake et~al. \citeyear{Drake2009}), and MIR data from \emph{WISE} (Wright et al. \citeyear{Wright2010}) and the newly released \emph{Near-Earth Object WISE Reactivation} mission (NEOWISE-R; Mainzer et al. \citeyear{Mainzer2014} ). 
CRTS is one of the largest time domain optical surveys currently operating, covering~$\sim$33,000~deg$^2$ with a baseline of 8 years and $\sim$250 exposures per year for each target. The survey is performed using unfiltered light but calibrated to a $V$-band zeropoint. We rejected the data points with large uncertainties ($>$0.2mag) and spurious points (usual outliers). Then we binned these data using the median value.
The \emph{WISE} has surveyed the full sky 1.2 times in four infrared bands \emph{W1, W2, W3}, and \emph{W4} centered at 3.4, 4.6, 12, and 22~$\mu m$ from 2010 January to September, on which its cryogen used to cool the \emph{W3} and \emph{W4} instruments was exhausted. Afterward it was extended an additional four months using \emph{W1} and \emph{W2}, and then it was placed in hibernation on 2011 February 1. On 2013 October 3, it is reactivated and named \emph{NEOWISE-R}, using only \emph{W1} and \emph{W2} (Mainzer et al. \citeyear{Mainzer2014}). So there is a $\sim$3.5 year gap in both \emph{W1} and \emph{W2} band light curves. \emph{WISE} scans a full sky area every half year and thus yielded 6-7 times the observations for each object up to the most recent public catalog.
Firstly, we removed bad data points of low image quality ("qi\_fact"$<$1) with a small separation to South Atlantic Anomaly ("SAA"$<$5), and flagged moon masking ("moon\_mask"$=$1). Then, we grouped the data by every half year as in our previous work (Jiang et al. \citeyear{Jiang2012}, \citeyear{Jiang2016}) and binned the data using the median value. Besides that we also collected the Stripe 82 multi-epoch data (Abazajian et al. \citeyear{Abazajian2009}) to visually compare to the CRTS \emph{V}-band data.

We collect the CL AGNs reported in the literature as much as possible to investigate their MIR behavior. These sources are then screened according to following criteria: 

(1) Seyfert galaxies or quasars that changed from Type 1 to Type $\geq$1.8 or from Type$\geq$1.8 to Type 1. Immediate types, 1.2 or 1.5, are abandoned because they may be partially obscured, complicating the interpretation.
 
(2) Variability amplitude is larger than 0.4 mag at 10$\sigma$ significance in either the \emph{W1} or/and \emph{W2} bands. 

(3) No source contamination within 6". The angular resolution is 6".1 and 6".4 at \emph{W1} and \emph{W2} (Wright et al. \citeyear{Wright2010}). We check the SDSS image for each object to exclude source contamination.
   
Twenty-four CL AGNs that satisfied criterion (1) are listed in Table \ref{t1}, among which, 11 objects meet the requirement (2). Finally, 10 follow all of these criteria, which are named SDSS J002311.06+003517.5, SDSS J081319.34+460849.5, SDSS J090902.35+133019.4, SDSS J101152.98+544206.4, SDSS J102152.34+464515.6, SDSS J132457.29+480241.2, SDSS J155440.25+362952.0 (iPTF 16bco), SDSS J225240.37+010958.7 (hereafter: J002311, J081319, J090902, J101152, J102152, J132457, J155440 and J225240, respectively), Mrk~1018 and NGC~2617. 

\begin{figure*}
	\figurenum{2}
	%\epsscale{0.95}
	\plotone{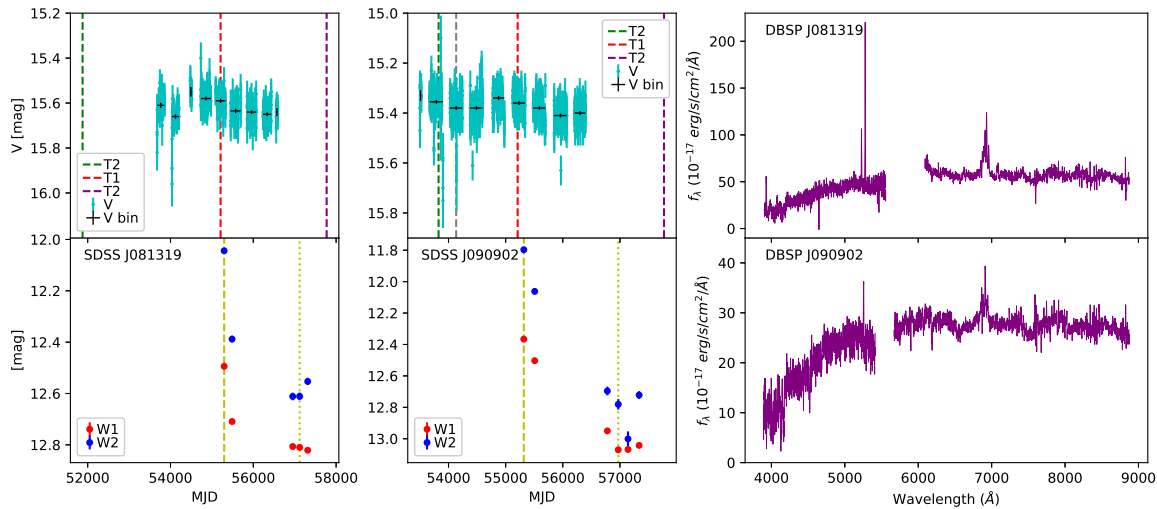}
	\caption{Light curves and spectra of J081319 and J090902. The purple dashed line marks the epoch of the DBSP spectrum, which confirms the second transition. Other symbols are the same as those in Fig \ref{fig:1}. The right panel presents the spectra obtained with DBSP (We used a 600/4000 grating for the blue side and a 316/7500 grating for the red side, and a D55 dichroic was selected. The slit was 1".5 and the exposure time was 10 minutes. The spectroscopic data were reduced following the IRAF standard routine.) }
	\label{fig:2}
\end{figure*}

\section{MID-INFRARED AND OPTICAL LIGHT CURVES}
{\label{lc}}
The 10 sources with significant MIR variability can be categorized into two equal subsamples
depending on whether the BELs appeared or disappeared, that is changed from Type 2 to Type 1 or vice versa. Below, we will introduce the MIR and optical light curves of the two classes, respectively. 

\subsection{From Type 1 to Type 2}{\label{f1t1}}
J101152, J102152, J132457, and Mrk~1018 are reported to experience transitions from Type 1 to Type $>$1.8 (MacLeod \citeyear{MacLeod2016}; McElroy et al. \citeyear{McElroy2016}; Runnoe et al. \citeyear{Runnoe2016}). Their CRTS and \emph{WISE} light curves are presented in Figures~\ref{fig:1}(a)-(d). Along with the transition, all the MIR light curves show apparent dimming ($>0.4$~mag) in both \emph{W1} and \emph{W2}. However, only J101152 and Mrk~1018 exhibit a similar dimming trend in the \emph{V}-band light curves. We have tried to fit their SDSS images with the PSF$+$Sersi$\acute{c}$ model using 2D decomposition software GALFIT (Peng et al. \citeyear{Peng2002}), assuming that the PSF and Sersi$\acute{c}$ represent the AGN and host galaxy emission, respectively. Our results suggest that the PSF component accounts for 26\% and 42\% of J102152 and J132457 in the SDSS \emph{r}-band, which is taken before the type transition. Due to seeing limit, the fitted PSF component can be considered as an upper limit of the AGN, which means that the real AGN fraction is even lower, and thus their optical variability is largely diluted. We have also noted that their $g$-band variability is much more significant ($\Delta g>1$~mag; MacLeod et al. \citeyear{MacLeod2016}) because it is less affected by the host galaxy. J101152 is totally AGN dominated and Mrk~1018 is a nearby ($z=0.035$) Seyfert galaxy with a well resolved nucleus, making their $V$-band variation are pretty detectable. We use yellow dashed and dotted lines to mark the significant/upper limit variation time scale of MIR bands of each source (see Figure \ref{fig:1})
 
Here we give some notes for each object.

(1) J101152: %Its V-band light curve dropped from a bright state to a long term dim state. Similar behavior also appears in its MIR-bands. 
 \emph{W1} faded by 1.10 mag within $\sim$4.00 years, while \emph{W2} faded by 1.76 mag.

(2) J102152: in $\sim$4.01 yr, W1 and W2 dimmed $\sim$0.59 and $\sim$0.62 mag respectively.  

(3) J132457: it showed the disappearance of BELs (MacLeod et al. \citeyear{MacLeod2016}), but recently showed the likely appearances of BELs (Ruan et al. \citeyear{Ruan2015}). Two MIR bands show similar re-brightened behavior. Both MIR bands dimmed continuously with \emph{W1} and \emph{W2} $\sim$0.47 and $\sim$0.45 mag respectively, in $\sim$4.53 years and then brightened after the turning point MJD$\simeq$57000. 

(4) Mrk 1018: it changed from Type 1.9 to Type 1 in 1984 (Cohen et al. \citeyear{Cohen1986}), but recently returned to Type 1.9 (McElroy et al. \citeyear{MacLeod2016}). In $\sim$4.49 years, \emph{W1} dimmed by 0.76 mag while W2 dimmed by 1.05 mag.

\subsection{From Type 2 to Type 1}
In FIgures \ref{fig:1}(e)-(h), we plot J225240, J002311, J155440, and NGC~2617, which show evidence of emerging BELs (Shappee et al. \citeyear{Shappee2014}; MacLead et al. \citeyear{MacLeod2016}; Gezari et al. \citeyear{Gezari2017}). Along with the transition, J155440 and NGC~2617 show a rising trend in their MIR bands. As for J225240 and J002311, the BOSS spectrum was taken during the \emph{WISE} epoch, which means their transitions happened before the \emph{WISE} survey (see panels (e) and (f)). So the \emph{WISE} and \emph{NEOWISE} missed the main uptrend transition period but covered the latter period. This situation is more obvious in J081319 and J090902 which are presented in Fig \ref{fig:2}. We note the six objects as follows.

(1) J225240: \emph{V}-band luminosity kept increasing ($\Delta$V$>$1 mag) during the transition, but decreased afterward, indicating it might change back to a dim state. \emph{W1} and \emph{W2} exhibit very similar variations. Because there's a gap between \emph{WISE} and \emph{NEOWISE}, we can only derive an upper limit of decreasing time scale as 3.52 years, with \emph{W1} and \emph{W2} dimming $\sim$0.65 and $\sim$0.52~mag respectively. 

(2) J002311: Optical/MIR variation behavior is very similar to that of J225240. The MIR bands dropped significantly in 1.51 years (from the gray dashed line to the yellow dotted line), with \emph{W1} and W2 dimming $\sim$0.41 and $\sim$0.27 mag respectively). Due to the gap in the MIR bands, we estimate the upper limit of decreasing time scale to be 4.51 yr (from the yellow dashed line to yellow dotted line).

(3) J155440: it was discovered as a transient on 2016 Jun 1 by iPTF (named as `iPTF 16bco', Gezari et al. \citeyear{Gezari2017}). From MJD=55409 to 57055, during $\sim$4.51 yr \emph{W1} and \emph{W2} brightened 0.69 and 0.94 mag respectively.

(4) NGC 2617: this was a Seyfert 1.8 galaxy in 2003 but showed appearance of BELs in 2013 (Shappee et al. \citeyear{Shappee2014}). From MJD=55506 to 56776, during $\sim$3.48 yr the \emph{W1} and \emph{W2} brightened 0.5 and 0.78 mag respectively.

\begin{figure}
	\figurenum{3}
	\epsscale{1}
	\plotone{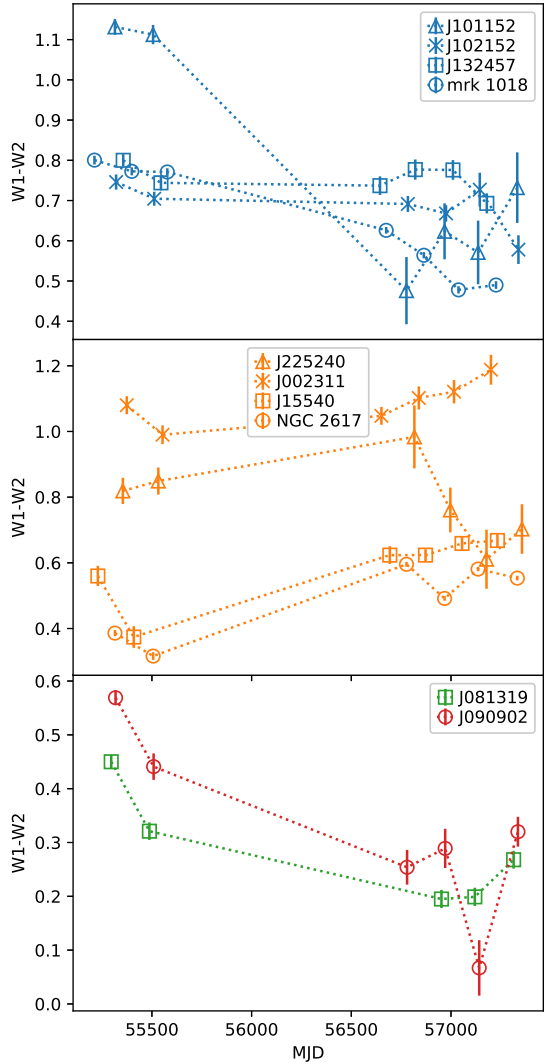}
	\caption{Color change of changing-look AGNs. Upper panel: changing from Type~1 to Type~2-like; middle panel: changing from Type~2-like to Type~1; bottom panel: J0813 and J0909.}
	\label{fig:3}
\end{figure}

(5) J081319 and J090902: J081319 was a Seyfet~1.8 galaxy in 2000, so was J090902 in 2006. Both of them were classified as Type~1 in 2010 (Runco et al. \citeyear{Runco2016}). Their MIR bands present a remarkable drop, in contrary to their previous transition, indicating a new change. We confirmed they have changed back to Type~1.9 using the Double Spectrograph (DBSP) of the Hale 200 inch telescope at Palomar Observatory. The spectrum of J081319 was obtained on 2017 January 8, while that of J090902 was obtained on January 18 (see Fig \ref{fig:3}). 

In summary, MIR bands of all the objects show a very similar variation to the optical data, which is in accordance with the type transition. In Fig \ref{fig:3}, we plot \emph{W1}-\emph{W2} of each sources. The sources with a transition from Type~1 to Type 2 likely change from AGN-like MIR-color (\emph{W1}-\emph{W2}$>$0.8) to galaxy-like MIR-color (\emph{W1}-\emph{W2}$<$0.5), vice versa (Stern et al. \citeyear{Stern2012}, Yan, Lin et al. \citeyear{YL2013}). We list the upper limits of MIR variation time as $\Delta T$ in column 9 and the change of MIR bands $\Delta \emph{W1}$ and $\Delta \emph{W2}$ in columns 11 and 12 of Table \ref{t2}.

\subsection{Remaining objects } {\label{R}}
Besides the 10 sources with significant MIR variability described above, we also examine the remaining targets with very weak/non-detectable MIR variations. For example, J233317.38-002303.4 and J214613.31+000930.8 which have the appearance of BELs, show significant uptrend in \emph{g}-band light curves but turn into a plateau with little variation after MJD$\sim$54500 (Macleod et al. \citeyear{MacLeod2016}), implying they might finish the transition. For Mrk~590, it has changed to Type 1.8$\sim$1.9 by 2006 (Denny et al. \citeyear{Denney2014}). Since their MIR light curves began at MJD$\sim$55179, \emph{WISE} likely missed their most significant transition epoch. For J233602.98+001728.7 and J022652.24-003916.5 their infrared radiations are too faint (\emph{W1}$\sim$15.4, \emph{W2}$\sim$14.8) to detect significant variation. The reasons above are applicable to most of the other objects not included in our sample, except J154507.53+170951.1, which is contaminated by another source within 6".

\section{Discussion}{\label{disc}}

\subsection{Physical scenarios}{\label{Phy_sc}}
The main motivation of this Letter is to explore the physical mechanism of CL AGNs using MIR variability. MIR emission at 3.4 and 4.6$\mu$m are mainly originated from hot-dust emission heated by AGNs (Netzer, \citeyear{Netzer2013bk}); hence the MIR emission should respond to the variation of accretion rate with a time delay. On the other hand, the MIR bands are not significantly affected by dust extinction; any detectable variability means a much larger amplitude of variability in the optical if CL is caused by the changing of obscuration. Even for J002311, whose MIR variation is the weakest ($\Delta$\emph{W2}=0.27) among the 10 objects in our sample, $\Delta$\emph{V}$\sim$6 is required assuming the extinction model of Fitzpatrick \& Massa (\citeyear{Fitzpatrick1999}). Such a dramatic optical variability is extremely rare for AGNs and does not agree with the optical light curves here.

Nevertheless, we further investigate the dynamical time scale of the obscuration. The size of the obscurer should be at least comparable to the torus to block the hot dust. With the inner radius of the torus  simply estimated from the dust sublimation radius (Netzer, \citeyear{Netzer2013bk}, 
$R_{sub}=0.5L_{46}^{0.5}(1800K/T_{sub})$\ pc), the crossing time for the obscuring material can be derived as $t_{cross}=0.073[r_{orb}/(lt-day)]^{3/2} M^{-0.5}_8arcsin(r/r_{orb})$ yr (LaMassa et al. \citeyear{LaMassa2015}), where $r_{orb}$ is the circular orbital radius of the obscurer, $M_8$ is the mass of black hole in units of $10^8M_{\odot}$, and $r$ is the true size of the obscured region (i.e. continuum emitting region or BLR size).
We adopted $r$ as the BLR size, which is estimated from calibrated $R-L$ relation (Bentz et al. \citeyear{Bentz2013}) and the calculated $R_{sub}$ and $t_{cross}$ are listed in Table \ref{t2} (columns 6 and 8). It can be seen clearly that $t_{cross}$ is much longer than the observed MIR variation time $\Delta T$ for all the 10 objects. 

Based on the analysis above, the CL behavior of our sample cannot be a result of the changes in obscuration. Once the obscuration case is ruled out, the MIR variability can be naturally attributed the hot-dust echo of the dramatic changing of the accretion rate. The time delay of the MIR and optical variability can offer us a unique opportunity to measure the radius of the torus.
Previous studies basing on $K$-band reverberation mapping suggested $R_{in}=R_{\tau_{K}}=0.47(L_{\rm bol}/10^{46}$\lum)$^{0.5}$ (Suganuma et al. \citeyear{Suganuma2006}).
Assuming $R_{\lambda}=(\lambda/\lambda_{K})^2R_{\tau_K}$, we got $R_{\rm \emph{W1}}=0.36$ pc ($0.11$ pc) and $R_{\rm \emph{W2}}=0.67$ pc ($0.21$ pc) for log$L_{bol}$ = 45~\lum (44~\lum), corresponding to time lag of 2.18 years (0.68 years) in the rest frame, or 1$\sim$3 years in the observed frame. We have tried to shift the MIR light curves backward a few of years, their variation pattern matching the optical ones well, giving fantastic evidence for the dust echo response to the accretion. In summary, we conclude that all of the 10 CL AGNs with significant MIR variability have undergone an drastic drop/rise in accretion rate.

\subsection{More information from the infrared variation}{\label{P}}

The MIR light curves can not only help us diagnose the physical scenarios of CL AGNs, but can also allow us to get more transition information and even predict new CL behaviors. Because MIR bands are AGN-heated hot-dust dominated, they are more efficient to detect when the AGN is optically weak comparable to its host galaxy (e.g. J102152 and J132457 in \S\ref{f1t1}). For J155440, it was proposed that the turn-on change state has occurred less than 500 days before 2016 June 1 (Gezeri et al. \citeyear{Gezari2017}). However, according to the \emph{V}-band and the MIR-band light curves, the turn-on event much likely began around 2013 June $\sim$ 2014 February (marked by purple dotted lines; see Figure \ref{fig:1}(g) ). The two objects J081319 and J090902 presented a very similar sign of new transition due to the MIR decline, and we confirmed they had changed back to Type~1.9. For J225240 and J002311, which changed from Type~2-like to Type~1, show a dramatic decline in MIR and optical light curves, suggesting they might also have changed back to type~2. We are planing to perform follow-up spectral observations to confirm of our conjecture.

\begin{table*}
	\caption{Properties of Changing-look AGNs}
	\centering 
	\resizebox{\textwidth}{!}
	{\begin{tabular}{cccccccccccc}
			\toprule
			(1)&(2)&(3)&(4)&(5)&(6)&(7)&(8)&(9)&(10)&(11)&(12)\\
			Name & z & log$M_{BH}/M_{\astrosun}$ & log$L_{bol}$ & log$L_{5100}$ & $R_{sub}$ & $R_{BLR}$ &$t_{cross}$ & $\Delta T$ & $R_{torus}$ &$\Delta W1$&$\Delta W2$\\
			& & & ( erg s$^{-1}$) & (erg s$^{-1}$) & (pc) & (lt-day) & (yr) & (yr)& (pc)&(mag)&(mag)\\
			\midrule
			J002311.06+003517.5 & 0.422 & 9.23  &  45.480 & 44.513 & 0.275 & 63.16 & 28.95  &  4.51  &  0.69 & 0.41 & 0.27\\
			J101152.98+544206.4 & 0.246 & 7.78  &  45.117 & 44.150 & 0.181 & 40.45 & 69.96   & 4.00  &   -   & 1.10 & 1.76\\
			J102152.34+464515.6 & 0.204 & 8.33  &  45.121 & 44.154 & 0.182 & 40.65 & 36.16   & 4.01  &  0.81 & 0.59 & 0.62\\
			J132457.29+480241.2 & 0.272 & 8.51  &  45.303 & 44.336 & 0.224 & 50.83 & 43.11    & 4.53  &   -   & 0.45 & 0.47\\
			J155440.25+362952.0 & 0.237 & 8$^{*}$ & 45.146$^{*}$ & 44.23$^{*}$ & 0.187 & 44.63 & 60.61  & 4.51 & - & 0.66 & 0.94\\
			J225240.37+010958.7 & 0.534 & 8.88  &  45.318 & 44.352 & 0.228 &   51.83  & 34.93   & 3.52  &   -   & 0.65 & 0.52\\
			Mrk 1018  & 0.035 & 7.4$\sim$7.9$^{*}$ & 44.491$^{*}$ & - & 0.088 & 24$^{*}$ & 21.0$\sim$37.4 & 4.49 & 0.82 & 0.76 & 1.05 \\
			NGC 2617  & 0.00142 & 7.6$^{*}$ &  44.03$^{*}$ & 43.12$^{*}$ & 0.05 & 11.42 & 10.4 & 3.48 & -    & 0.50 & 0.78 \\
			J081319.34+460849.5 &    0.054  &  6.98$\sim$7.28$^{\star}$   &  43.56$\sim$44.01$^{\star}$  & 42.65$\sim$43.10$^{\star}$  &  0.03$\sim$0.05    &  6.44$\sim$11.14  &  9.57$\sim$15.34  &  5.00  &  -  &  0.32  &  0.57  \\
			J090902.35+133019.4 &   0.050  &  7.03$\sim$7.32$^{\star}$   &  43.42$\sim$43.87$^{\star}$  &  42.51$\sim$42.96$^{\star}$  &  0.03$\sim$0.04    &   5.43$\sim$9.39  &  7.06$\sim$11.31  &  4.53  &  -  &  0.70  &  0.98  \\
			\bottomrule
	\end{tabular}}
	\tablecomments{ The table lists redshift, black hole mass, bolometric luminosity, monochromatic luminosities at 5100$\AA$ and sublimation radius $R_{sub}$ of each source in 2$\sim$6 columns. The estimated characteristic radius of BLR and crossing time scale $t_{cross}$(in observation frame) for obscuration, along with the observed MIR variation time scale $\Delta T$ are listed in 7$\sim$9 columns. The column 10 lists the estimated upper limit of the radius of torus. Column 11 and 12 list the change of MIR bands. The luminosities, redshift and $M_{BH}$ are taken from Shen et al. \citeyear{Shen2011}, except the data with a asterisk annotation are from the corresponding paper of which the source is reported, and the data with a star annotation are estimated from H$\alpha$ (Greene et al. \citeyear{Greene2005}) by fitting the SDSS spectrum. }
	\label{t2}
\end{table*}
\section{Conclusion}\label{C}
We investigate the 10 reported CL AGNs confirmed in the optical spectrum in the literature. Combining with \emph{WISE} and \emph{NEOWISE} multi-epoch photometric data (\emph{W1}, \emph{W2}), all the 10 sources have obvious MIR variation ($>$0.4 mag) at level $>10\sigma$. The obscurer passing across the line of sight could not cause such a large variation, due to the extinction and dynamical obscuration timescale that failed to support it. Four sources with disappearance of BELs, namely, J101152, J102152, J132457, and Mrk 1018, show a similar significant decline ($>10\sigma$) in MIR light curves which echo to the optical variation. We suggest that their CL is owing to a drop in the accretion rate. Among the four sources showing emerging BELs, two objects J155440 and NGC 2617, feature a remarkable increase in MIR luminosity, in accordance with their transition and also the increasing \emph{V}-band tendency. We suggest their CL is due to accretion rate rising up. Two other object, J225240 and J002311, abnormally display a significant decrease in MIR bands, indicating they might undergo a second transition and change back to the previous Type 2, which should be confirmed by follow-up observations. In particular, J081319 and J090902, which are reported to have changed from Type~1.8 to Type~1, have a significant ($>10\sigma$) decrease in MIR signals. We confirmed that they changed back to Type~2. Further and repeated spectroscopic monitoring of sources with large MIR variability sources could be worthwhile.

\section*{Acknowledgments}
We acknowledge the anonymous referee for valuable comments that helped to improve the Letter. And we acknowledge Andrew Drake for providing the unreleased CRTS data (Drake et al. \citeyear{Drake2009}) of J155440. Also, we acknowledge Jordan N. Runco for helpful information on J081319 (Runco et al. \citeyear{Runco2016}). We thank Luming Sun for discussion and Zhihao Zhong for his perfect spectrum reduction skill. This project is supported by National Basic Research Program of China (grant No. 2015CB857005), the NSFC through NSFC-11233002, NSFC-11421303, NSFC-116203021, and U1431229, jointly supported by the Chinese Academy of Science and NSF. This research has made use of the NASA/IPAC Infrared Science Archive, which is operated by the Jet Propulsion Laboratory, California Institute of Technology, under contract with the National Aeronautics and Space Administration. The CSS survey is funded by the National Aeronautics and Space Administration under grant No. NNG05GF22G issued through the Science
Mission Directorate Near-Earth Objects Observations Program.  The CRTS survey is supported by the U.S.National Science Foundation under grants No. AST-0909182. This research made use of Astropy, a community-developed core Python package for Astronomy (Astropy Collaboration, \citeyear{Astropy2013}).

\end{document}